\documentclass[aps,twocolumn,nofootinbib]{revtex4}
\usepackage{dcolumn} %Align table columns on decimal point
\usepackage{bm} %Load bold math symbols
\usepackage{amsmath}
\usepackage{txfonts}
\usepackage{stmaryrd}
\usepackage{mathrsfs}
\usepackage{multirow} 
\usepackage{graphicx}
\usepackage{amssymb}

\begin{document}

\title{Inflation from a Non-Local Theory of Gravity}
\author{Philip Stephens}
\affiliation{Physics Department, Lancaster University, Lancaster LA1 4YB}

\begin {abstract}
In this paper, I have used direct methods in order to evaluate the inflationary dynamics of $f(R,\square R)$ gravity with Lagrangian $f(R, \square R) = R + c_{0} R^{2} + \sum_{i=1}^{\infty}c_{i}R \square^{i} R$. This is found to have equivalent solutions to $R+R^2$ gravity, and to be dynamically equivalent to chaotic inflation. Perturbation spectra are calculated, and fitted to the COBE normalisation. This is found to correspond to a large field inflation model. 
\end {abstract}
\maketitle

\section{Introduction}
Ever since Einstein and Hilbert first noticed that Einstein's Field equations could be derived from an action principle, there has been interest in the effects of inserting additional terms into the Einstein-Hilbert action. However, it was the rise of the inflationary paradigm \cite{guth}\cite{starobinsky}\cite{linde}, coupled with the discovery of field theory, that has provided motivation for a more widespread investigation of this field. The first and foremost motivation is that we expect that gravity should have a field theoretic description, in which case the Einstein-Hilbert action should be treated as some kind of effective field theory. In this case we expect that there should be present a very general class of terms with higher powers of the curvature tensor, that are suppressed according to some mass scale. It is hoped that this will also make gravity into a renormalisable theory at low energy. These terms can have significant effects either at early times when the universe has higher curvature, or at very large scales where tiny effects can become cumulatively significant, at least in areas where the mass density is sufficiently low. It is generally thought that additions based around the Ricci Scalar should be effective toy models for more general tensor additions.

Secondly, the question of finding a theoretically well motivated construction for inflation is still open. Within the slow roll approximation, scalar fields can be fitted to the data with an arbitrary degree of precision, fitting one into a well motivated particle physics construction has proved to be much more difficult. Examples exist in the Minimal Supersymmetric Standard Model \cite{MSSM1}\cite{MSSM2}\cite{MSSM3}, or the Higgs Inflation Model \cite{Higgs}, but until experimental verification, they remain only theories. Until such questions are answered, positing the existence of a slowly rolling scalar field seems no more likely (and in many ways less so) than the idea that gravity could itself drive inflation\cite {Khoury}.

The aim of this paper is to investigate the inflationary dynamics of adding a series of arbitrarily high derivatives of the form $R\square^{k}R$. The addition of an infinite series of such terms was previously found to be ghost free, and to give rise to an asymptotically free theory of gravity, and a bouncing cosmology \cite{Anupam}.

\section{Deriving the evolution of the background}
In this paper we consider a Lagrangian given by
\begin{equation}
\label{L}
L = f(R,\square R)\sqrt{-g},
\end{equation}
where $f(R)$ is given by
\begin{equation}
f(R)=R + c_{0} R^{2} + \sum_{i=1}^{\infty}c_{i}R \square^{i} R,
\end{equation}
and $\square$ has its usual meaning as the d'Alembertian. We use here a result of Schmidt \cite{schmidt} for the Lagrangian (\ref{L}) giving
\begin{align}
\frac{1}{\sqrt{-g}}\frac{ \delta L}{\delta g_{ij}} = & GR^{ij} - \frac{fg^{ij}}{2} - G^{;ij} + g^{ij}\square G  \nonumber \\ & + \sum_{A=1}^{\infty}\frac{g^{ij}}{2}(F_A (\square^{A-1} R)^{;k})_{;k}  -  \sum_{A=1}^{\infty}F_{A}^{;(i}(\square^{A-1}R)^{;j)},
\end{align}
where we have defined the following functions:
\begin{equation}
F_{A} = \sum_{j=A}^{k} \square^{j-A} \frac{\partial f}{\partial \square^{j} R}, \qquad G = F_{0}.
\end{equation}

If we assume that the dimension of our space time is 4, then we can take the trace of the variational derivative to find

\begin{equation}
\label{trace}
3 \square G = 2f-GR-2\sum_{A=1}^{\infty} F_{A} \square^{A} R + 2\sum_{A=1}^{\infty} F_{A;i}( \square^{A-1} R)^{;i}.
\end{equation}

As is usual in a homogenous isotropic universe, only the 00 component of the field equations and the trace provide independent constraints, and it will be sufficient to work with the much less complicated trace equation to establish the evolution of the curvature scalar. 

\subsection{Finding solutions}
Before inserting our Lagrangian into the trace equation, it is instructive to input two simpler Lagrangians defined by

\begin{eqnarray}
f(R, \square R ) = R + \alpha R^2, \\
f(R, \square R ) = R + \alpha R^2 + R\square R.
\end{eqnarray}

In the first case the trace equation is reduced to 
\begin{equation}
3 \square G = 2f - GR,
\end{equation}
with
\begin{eqnarray}
G = \frac{\partial f}{\partial R} = 1 + 2\alpha R,
\end{eqnarray}

and consequently we have the equation of motion for R as
\begin{equation}
\label{R}
\square R = \frac{1}{6\alpha} R.
\end{equation}
Under the assumptions of homogeneity and isotropy the d'Alembertian reduces to a double time derivative (clearly under such conditions all the space derivatives must be zero), in which case it is clear that R has the solution
\begin{equation}
 R = A\exp(\pm\frac{t}{\sqrt{6\alpha}}).
\end{equation} 
It is interesting that the sign of alpha gives the difference between oscillatory and exponential behaviour. Notice also that $R=0$ is a solution, that is, a flat space time remains a stable solution in the absence of matter, and indeed, is an attractor of the vacuum solution. The equation of motion does not specify whether the curvature is increasing or decreasing, we wish to study inflationary modes, and therefore choose that $\alpha > 0 $ and $\dot{R} < 0$. Using again the assumptions of homogeneity and isotropy we are left with a Robertson-Walker space time, and this implies that $R = 12H^{2}$ \cite{misner}. With (\ref{R}) given above, and the usual proscription for H in terms of the scale factor, it is clear that this relation can be integrated to yield that
\begin{equation}
\ln a(t) = -A \sqrt{2\alpha} \exp(-\frac{t}{\sqrt{24\alpha}}) + B, 
\end{equation}
yielding
\begin{equation}
\label{scale}
a(t)= B'\exp( - A \sqrt{2\alpha} \exp(-\frac{t}{\sqrt{24\alpha}})),
\end{equation}
where B/B$'$ have been introduced as constants of integration. 

Let us continue with the calculation of the background evolution of 
\begin{equation}
f(R, \square R ) = R + \alpha R^2 + \gamma R\square R.
\end{equation}
Using the trace of the field equation (\ref{trace}), we find that
\begin{equation}
6\alpha\square R + 6 \gamma \square^{2}R = R - 2\gamma R\square R + 2\gamma \triangledown_{i}R\triangledown^{i}R,
\end{equation}
and under the conditions of homogeneity and isotropy we can make the identification that $\triangledown = \frac{\partial}{\partial t}$, leading to
\begin{equation}
2\gamma\left( R\frac{\partial^{2}R}{\partial t^{2}} - \left(\frac{\partial R}{\partial t}\right)^{2}\right) = R - 6\alpha \frac{\partial^{2}R}{\partial t^{2}} - 6 \gamma \frac{\partial^{4}R}{\partial t^{4}}.
\end{equation}
One can immediately see that for any exponential solution the left hand side vanishes identically, so let us put $R = \exp{\beta t}$ into the right hand side to produce,
\begin{equation}
0 = 6\gamma\beta^{4} + 6\alpha\beta^{2} - 1,
\end{equation}
which clearly has solutions if
\begin{equation}
\beta^{2} = \frac{-6\alpha \pm \sqrt{36\alpha^{2} + 24\gamma}}{12\gamma}.
\end{equation}
This clearly has real solutions for $\beta$ if $\gamma>0$, or if $\gamma < 0$ but $|\gamma|<\frac{3}{2}\alpha^{2}$. Given that we expect each term with higher derivatives of the metric to be associated with higher powers of some mass scale M, combined with some coupling constant of order 1, we find that the second of these conditions is nearly automatically satisfied, as it gives 
\begin{equation}
|\lambda_{\gamma}| < \frac{3}{2}\lambda_{\alpha}^{2}.
\end{equation} 
Thus we see that despite the extra terms the solutions are nearly identical, still yielding exponential behaviour of R, with the corresponding form for the scale factor Eq.(\ref{scale}). There is a pleasing symmetry about this, that extra terms do not disrupt the form of the equations. 

\subsection{Solutions to the full infinite derivative Lagrangian}
In this section we present results based on a direct analysis of 
\begin{equation}
f(R)=R + c_{0} R^{2} + \sum_{i=1}^{\infty}c_{i}R \square^{i} R,
\end{equation}
and we need to insert this into the trace of the field equation Eq.(\ref{trace}). This is a slightly tedious calculation made easier by noting $2F - GR = R$ is an easily established identity. In order to see the nature of the solutions it is convenient to group the terms according to whether they are linear or quadratic in R, giving
\begin{equation}
3\square G - R = \sum_{A\geq 1}\left( \triangledown_{i}F_{A}\triangledown^{i}\square^{A-1}R - F_{A}\square^{A}R\right).
\end{equation}
It is easy to see that an exponential solution can only solve this if one of the sides of this equation reduces to zero. If we make the assumption of homogeneity, so that we can drop the space derivatives, then we find immediately that the right hand side will reduce to zero for any exponential solution. In order to understand the solutions physically, we shall make explicit the mass dependence at this time, by writing,
\begin{equation}
R = A \exp(-bMt)
\end{equation}
and, 
\begin{equation}
c_{i} = \frac{\lambda_{i}}{M^{2i+2}}.
\end{equation}
Using these, we are left with
\begin{equation}
0 = 3\square \left(1 + 2\frac{\lambda_{0}R}{M^{2}} + 2\sum_{i=1}^{\infty} \frac{\lambda_{i}}{M^{2+2i}}\square^{i}R\right) - R,
\end{equation}
reducing to
\begin{equation}
0 = 6\sum_{i=0}^{\infty}\lambda_{i} b^{2i+2} - 1,
\end{equation}
upon insertion of the trial solution. This type of polynomial will have (at least one) real solution for $b$ under some very general constraints. In particular  if we consider that all the $\lambda_{i}$ are about 1, and then this is a becomes a simple geometric series in terms of b. In particular we can use the geometric sum noting that,
\begin{equation}
\frac{1}{6} = \frac{b^2}{1-b^2},
\end{equation}
giving $b=\frac{1}{\sqrt{7}}$.
We had at this point a free choice of sign under the square root, but we elected for the positive sign given that we are looking for inflationary solutions. At any rate, negative solutions will be unstable as the curvature will tend exponentially towards a singularity. In practice the exact value of b is not necessary, given that we shall treat the mass scale $M$ as a free parameter to be determined by the COBE normalisation.

\section{Inflationary dynamics.}
In this section we shall investigate the inflationary dynamics, in particular we shall find an exact form for the scale factor, and show that inflation will end naturally. We will then show that the the theory is slow rolling, and that it shares a dynamical equivalence with a scalar field theory, which shall be used to evaluate the spectrum of density perturbations in the following section. 

Firstly, we note that under the assumptions of homogeneity and isotropy, the Freidmann-Robertson-Walker metric is necessarily a solution to any metric theory of gravity, and thus we can make use of the well known relation (see for example \cite{Weinberg}) $R = 12H^{2}$. It is not difficult to integrate this to find that
\begin{equation}
a(t) = a_{0}\exp\left(\frac{-2}{bM}\sqrt{\frac{R_{0}}{12}}\exp\left(\frac{-bMt}{2}\right)\right).
\end{equation}
The subscript $0$ indicates that the quantity is defined at $t=0$, which does not necessarily correspond to the beginning of the universe, but to an arbitrary time when we choose to define both $R_{0}$ and $a_{0}$. We shall use that $R_{0} = M_{Pl}^{2}$ as this seems to be a natural scale for $R$ at the start of the inflationary era. We also note that despite the odd form of a, it is a quasi de-Sitter space, at least initially, as provided $\frac{-bMt}{2}<<1$, then $\exp(\frac{-bMt}{2}) \approx 1 - \frac{-bMt}{2}$. With this exact solution known, we can investigate the behaviour of $\ddot{a}(t)$, and we find that
\begin{equation}
\ddot{a}(t) = a\sqrt{\frac{R_{0}}{12}}\left\{\sqrt{\frac{R_{0}}{12}}\exp\left(-\frac{bMt}{2}\right) - \frac{bM}{2}\right\}\exp\left(-\frac{bMt}{2}\right)
\end{equation}
is monotonically decreasing, and reaches $\ddot{a}=0$, corresponding to the end of inflation, at a finite time given by
\begin{equation}
t_{end} = \frac{2}{bM}\ln \left(\frac{2}{bM}\sqrt{\frac{R_{0}}{12}}\right).
\end{equation}
If we look at the parameter $\frac{\dot{H}}{H^{2}}$ then we find that it is less than 1 until exactly $t_{end}$. Physically, this parameter is equivalent to the slow roll parameter of a scalar field theory, (see, for example, \cite{lythe}). This should come as no surprise, as in this theory the extra terms of the Lagrangian all reduce to the form of an $R^2$ term during the inflationary scenario, when homogeneity and isotropy can be invoked, this indicates that our theory has a single extra (scalar) degree of freedom compared to standard Einstein gravity. It is then no surprise to find a dynamical equivalence to Einstein Gravity plus one scalar field. This equivalence has also been extensively investigated in the form of a conformal transform\footnote{I believe that conformal transformations may obscure physically relevant similarities between theories. In particular it is not always clear what approximations made in the conformal frame correspond to, when a conformal transformation represents a point-wise scaling of such physically meaningful quantities as mass-energy and length. The mathematical equivalence of two frames related by a conformal transformation is only preserved under the redefinition of units in the conformal frame to account for the point wise rescaling of length, see for example, \cite{pseudo}}. In order to show the equivalence of our theory with scalar field theories, we shall use the slow roll approximations to define an implicit field $\phi$, which we shall then show fulfils the canonical scalar field equation, justifying the use of the slow roll predictions.  As in the use of the conformal transformation, the field $\phi$ is a purely mathematical convenience - in practice it is a substitution for $R$, reducing our field equations to the form of a scalar field. However, unlike a conformal transform, there is no mixing of degrees of freedom. 

In essence we use the relations,
\begin{eqnarray}
3M_{Pl}^{2} H^{2} =  V(\phi(t)) \\
3H\dot{\phi} = - \frac{dV(\phi(t))}{d\phi}\\
\end{eqnarray}
leading to
\begin{eqnarray}
\phi = \frac{4M_{Pl}}{\sqrt{bM}} \sqrt[4]{\frac{R_{0}}{12}} \exp \left(-\frac{bmt}{4}\right), \\
V(\phi) = \frac{3b^{2}M^{2}}{16^{2}M_{Pl}^{2}}\phi^{4}. \label{V}
\end{eqnarray}
This method of direct equivalence should be accurate, provided that $\phi$ fulfils the canonical wave equation to within a good approximation, that is, $\ddot{\phi}<< 3H\dot{\phi}$. It is easy to check that this is certainly satisfied provided that $\phi$ is greater than around $5M_{Pl}$. This is important, as it ensures that the perturbations will have the same classical evolution as in the slow roll paradigm, when cosmologically significant scales move outside the horizon. It is generally believed that the effects of a truly quantised gravity will be sub-dominant to the effects of the classical evolution. Thus this extended gravity theory will have cosmological predictions that are necessarily similar to the slow roll paradigm. We also note that the form of the potential is essentially that of chaotic inflation, but the parameter $\lambda$ is now seen to be a function of the mass scale $M$, potentially explaining the otherwise unnatural smallness of the coupling constant. 

\section{Computation of the Slow roll predictions}
Firstly we begin with the slow roll parameters,
\begin{eqnarray}
\epsilon = \frac{M_{Pl}^{2}}{2} \left(\frac{V'}{V}\right)^{2} = \frac{8M_{Pl}^{2}}{\phi^{2}} \\
\eta = M_{Pl}^{2}\frac{V''}{V} = \frac{12M_{Pl}^{2}}{\phi^{2}}.
\end{eqnarray}
Thus we can see that inflation comes to an end at $\phi = 2\sqrt{2}M_{Pl}$, which is precisely $t_{end}$, given above. The next important result is the total number of e-folds, this is given by
\begin{eqnarray}
&N = \ln\frac{a(t_{end})}{a(t_{start})} \\
&= (\frac{2}{bM}\sqrt{\frac{R_{0}}{12}}\left(\exp(\frac{-bM}{2}t_{start} - \exp(\frac{-bM}{2}t_{end})\right). \\
&=\frac{1}{8M_{Pl}^{2}}\left(\phi^{2}_{start} - \phi_{end}^{2}\right)
\end{eqnarray}

Thus the dominating factor in the number of e-folds is where inflation starts, rather than when it ends, although this is not altogether surprising given that inflation effectively ends at a fixed time controlled only by the parameter $M$. In practice the choice of $t_{start}$ is implicitly tied to the choice of $R_{0}$, it is convenient therefore to choose $t=0$ as the start time. Given that it is expected that $R_{0}$ will be of the order $M_{Pl}^{2}$ and that M will be several orders of magnitude smaller, it can be seen that this theory produces more than enough inflation to account for observation. We will check the value of $M$ against the COBE normalisation next. We use the simplifying assumption of \cite{lythe} that cosmologically interesting scales cross the horizon when $N_{remaining} = 50,$\footnote{This value corresponds to the generally accepted number for when observable scales passed outside the Horizon, but it depends on the assumption of Einstein gravity. Even then it is only an approximation and so it is possible to increase this number as high as 60 without any serious constraints. Further, given that the behaviour of H becomes non canonical a few e-folds before the end of inflation, and until reheating produces sufficient radiation to enter radiation domination, the faster than expected (cf radiation domination) decrease in H should could leave us with a higher than expected N, and could favour N=60.} corresponding to $\phi = 20.2M_{Pl}$. We continue the paradigm by evaluating the spectral index $n$ and its running, evaluated at $N_{50}$. (All data for reference has been taken from \cite{WMAP}).
\begin{eqnarray}
& n = 1 + -6\epsilon + 2 \eta = 0.941 \\
& \frac{dn}{d\ln K} = 5.77 \times 10^{-4}.
\end{eqnarray}
We are now in a position to evaluate the COBE normalisation, which is given by the expression
\begin{equation}
\delta^{2}_{H}(k) = \frac{1}{150\pi^{2}M_{Pl}^{4}} \frac{V}{\epsilon}.
\end{equation}
As usual this is evaluated at the pivot scale, corresponding to around 50-e folds before the end of inflation, giving the expression
\begin{equation}
\delta^{2}_{H}(k) = 67.2 \times \frac{b^{2}M^{2}}{M_{Pl}^{2}},
\end{equation}
which, using $b = \frac{1}{\sqrt{7}}$ as calculated in section 1, and the standard COBE normalisation we find that $M = 6.16 \times 10^{-6} M_{Pl} = 1.5 \times 10^{13}GeV/c^{2}$. This is in keeping with results obtained via the conformal transformation, see for example, \cite{mukhanov}. The final piece of analysis concerns the scalar to tensor ratio, given by
\begin{equation}
r=12.4\epsilon_{*} = 0.24.
\end{equation}
In reality, this is a little high, the WMAP data strongly favouring $r<0.2$. However, as alluded to in an earlier foot note, this is not actually an insurmountable obstacle, as the choice of $N_{remaining}$ is based on the thermal history, with many authors preferring 55 rather than 50, and some textbooks suggest that values as high as 70 are possible, if disfavoured. It is perhaps sensible to think that (\ref{scale}) may govern the evolution of H for around one Hubble time before the radiation density has grown enough to dominate, this would tend to increase $N_{*}$ by a few e-folds. However, a choice of 70 is needed to bring the results into the 2-$\sigma$ region of the WMAP probe, and that is probably too high, unless a theory of reheating should cast the standard thermal history of the universe in doubt. 

We have not discussed reheating at all in this model. The scalar field above does not correspond to a real scalar field, and will not oscillate around a minimum as in the standard scalar field examples. However, at the point where accelerated expansion ends, there is still a large energy density locked in the Ricci scalar, and if this couples to some particle physics model, as in the Higgs \cite{Higgs} inflation model, then this could decay into the particle sector. Alternatively, the particle physics could be wholly decoupled from the gravity action, and contain fields of its own which can decay into the standard model. 

\section{Conclusion}

We have shown, via a direct evaluation, that our infinite derivative Lagrangian has the same inflationary solutions as is found in $R + R^{2}$ gravity, and that these solutions are functionally identical to that of a scalar field with a $\phi$-fourth potential when cosmologically interesting scales pass outside the horizon. This author eschewed the use of a conformal transformation in this example, seeing it as an unnecessary complication, and notes that here the use of a direct method has revealed the phenomenological equivalence with a phi fourth scalar field which is obscured (though presumably it still exists - see \cite{pseudo}) by a conformal transformation.  This therefore has similar inflationary dynamics to chaotic inflation, only now the small value of the dimensionless coupling constant lambda (identified with the constants in Eq.(\ref{V})) needed to match the COBE normalisation is seen to be motivated as it is found to be dependent on the mass scale $M$. Unfortunately, it therefore suffers the same fate as phi-fourth inflation, which is generally thought to be outside the 2 sigma parameter space of the 5 year WMAP data.

\end{document}